\documentclass[12pt]{iopart}

\usepackage{amsfonts,amssymb}
\usepackage{graphicx}
\usepackage{epstopdf}
\usepackage{subfig}
\usepackage{booktabs}
\usepackage[subfigure]{tocloft}
\usepackage{setspace}
\usepackage{tocloft}
\usepackage{array}
\usepackage{url}
\usepackage{bm}
\usepackage{multicol}
\usepackage{harvard}
\usepackage{iopams}
\graphicspath{{images/}}
\begin{document}

\title{Robustness of Optimal Energy Thresholds in Photon-counting Spectral CT}

\author{Yifan Zheng$^{1,2,5}$, Moa Yveborg$^3$, Fredrik Gr{\"o}nberg$^1$, Cheng Xu$^{1,3}$, Qianqian Su$^2$, Mats Danielsson$^{1,3}$ and Mats Persson$^{4}$}

\address{$^1$ Department of Physics, Royal Institute of Technology, Albanova University Center, SE-106 91 Stockholm, Sweden}
\address{$^2$ Key Laboratory of Particle and Radiation Imaging (Ministry of Education) and Department of Engineering Physics, Tsinghua University, Beijing 100084, China}
\address{$^3$ Prismatic Sensors AB, Roslagstullsbacken 21, Stockholm 106 91, Sweden}
\address{$^4$ Departments of Bioengineering and Radiology, Stanford University, Stanford 94305, California, USA}
\address{$^5$ Author to whom any correspondence should be addressed}
\ead{yifanzhe@kth.se}
\vspace{10pt}
\begin{indented}
	\item[]August 2018
\end{indented}

\begin{abstract}
An important question when developing photon-counting detectors for computed tomography is how to select energy thresholds. In this work thresholds are optimized by maximizing signal-difference-to-noise ratio squared (SDNR$^2$) in an optimally weighted image and signal-to-noise ratio squared (SNR$^2$) in a gadolinium basis image in a silicon-strip detector and a cadmium zinc telluride (CZT) detector, factoring in pileup and imperfect energy response in both detectors. To investigate to what extent one single set of thresholds could be applied in various imaging tasks, the robustness of optimal thresholds with 2 to 8 bins is examined with the variation of phantom thicknesses and target materials. In contrast to previous studies, the optimal threshold locations don't always increase with increasing attenuation if pileup is included. Optimizing the thresholds for a 30 cm phantom yields near-optimal SDNR$^2$ or SNR$^2$ regardless of target tissue types and surrounding attenuation for both detectors. Having more than 3 bins reduces the need for changing the thresholds depending on anatomies and tissues. Using around 6 bins or 8 bins may give near-optimal SDNR$^2$ or SNR$^2$ without generating an unnecessarily large amount of data.
\end{abstract}


\vspace{2pc}

\submitto{\PMB}

\begin{spacing}{1}

\section{Introduction}
\label{sect:intro}  
Photon-counting spectral CT with multiple energy bins has several advantages over conventional energy-integrating CT \cite{Si2017Review,Gutjahr2016Human}, including better image quality \cite{schmidt2009optimal,Cormode2017Multicolor}, lower radiation dose \cite{chen2015optimization,le2010radiation,Symons2017Low}, and K-edge imaging feasibility \cite{roessl2007k,Symons2017Photon}. The signal-difference-to-noise ratio (SDNR) can be increased by up to 60\% in photon-counting CT with the same radiation dose as conventional CT \cite{bornefalk2010photon}. K-edge imaging can help to better distinguish and quantitate contrast agents \cite{schlomka2008experimental,roessl2007k}.

Photon-counting spectral CT detects incident photons and sorts them into specific energy bins according to the deposited energy. Conventionally, the first threshold is set higher than the noise floor of the detector and the other thresholds are placed either in uniform increment or on some certain values \cite{persson2014energy,roessl2007k}. The placement of energy thresholds is important in improving image quality \cite{mullner2015feasibility}. To yield the maximum SDNR$^2$ or to minimize the Cram{\' e}r-Rao lower bound (CRLB) in K-edge imaging \cite{Roessl2009Cram}, and to reduce patient dose at the same time, the number and position of energy thresholds have to be optimized. A greater number of energy bins will benefit maximizing SDNR$^2$ or minimizing CRLB but will generate an unnecessarily large amount of data at the same time. Thus the tradeoff between better SDNR$^2$ or CRLB and less data must be considered to optimize the number of bins. Moreover, the thresholds are set before exposure and it is inconvenient to change them during the scan. Although the thresholds could be changed or adjusted based on patient size for each scan but this will increase the engineering complexity. Also, for each patient, the phantom thickness is different along different projection rays and varies with rotation angle, so that changing the thresholds for various patients may not give an optimal performance at a specific projection. So it is desirable to find one single set of thresholds that gives close to the optimal performance regardless of the variation of patient attenuation and tissue compositions. In previous studies, the placement of optimal thresholds was calculated \cite{wang2011pulse,wang2009optimal,wang2011sufficient} and the optimal threshold energies were found to increase with phantom thicknesses as a result of the beam hardening effect \cite{shikhaliev2009projection,wang2009optimal,roessl2006optimal,roessl2011sensitivity}. Better image quality was yielded with some thresholds placed around the K-edge of the contrast agent \cite{cormode2010atherosclerotic,mullner2015feasibility}. When optimzing energy thresholds in a cadmium zinc telluride (CZT) detector with six bins, the 'close bracketing' effect was found where two optimal thresholds closely surround the K-edge of the contrast agent \cite{roessl2011sensitivity}. However, previous studies have not investigated to what extent one single set of energy thresholds could be applied in various imaging tasks; and the detectors in those studies were assumed to be ideal, neglecting pulse pileup, charge sharing or K-fluorescence. A precise detector response plays an important role in setting optimal thresholds in real applications \cite{roessl2011sensitivity}.

The purpose of this work is to investigate the robustness of optimal energy thresholds with respect to the variation of phantom attenuation and tissue types. The optimal energy thresholds are examined based on two types of photon-counting detectors: silicon-strip detectors and CZT detectors. The influences of pulse pileup on the optimal threshold placement and the robustness of optimal thresholds are also analyzed in both detector models. In Sec. \ref{sect:MM}, simulation setup and the optimization algorithm are described, with SDNR$^2$ in an optimally weighted image and signal-to-noise ratio squared (SNR$^2$) in a gadolinium basis image used as the figures of merit (FOM) respectively to evaluate threshold optimization. The results are reviewed in sections \ref{sect:result} and \ref{sect:discussion}.

\section{Materials and Methods}
\label{sect:MM}
The placement of energy thresholds is optimized for a silicon-strip detector and a CZT detector separately. The two studied detectors are two example detector systems selected for the purpose of the robustness of optimal thresholds in this study. They have different parameters based on real-world detector systems but the study should not be seen as a comparison between the two detectors. Such a comparison would require more detailed spectral response models and would have to consider a range of values for different parameters such as detector absorption thickness, pixel size etc. Therefore we pick the following parameters as an example study. The X-ray beam after 0.8 mm thick beryllium and 6 mm thick aluminum filtration passes through a D cm thick water phantom with an embedded 1 cm thick imaging target and reaches the detector. The source-to-detector distance is 1 m. The X-ray spectrum is generated by a tube with tungsten anode at 11 degrees. The tube voltage is set to 120 kilovoltage peak (kVp) and the tube current is 500 mA. The acquisition time for each projection is 100 $\mu$s \cite{prince2006medical}. The total fluence rate at the detector is 2.0 $\times$ 10$^6$ mm$^{-2}$s$^{-1}$mA$^{-1}$ \cite{persson2016upper}. Photon counts scattered by the object are not taken into consideration by assuming a perfect anti-scatter grid.

The silicon-strip detector has a maximum of 8 bins and is applied in edge-on geometry with one edge towards incident X-rays \cite{bornefalk2010photon,persson2014energy,xu2013energy,liu2014silicon,liu2015spectral,liu2016count}. It has a 0.5 mm thick dead layer on the front edge and a 3 cm thick active absorption layer subdivided into nine depth segments, with lengths exponentially increased to ensure a uniform count rate in each segment \cite{liu2015spectral}. The pixel size of each detector element is 0.4 $\times$ 0.5 mm$^2$ \cite{xu2013energy,liu2014silicon}. The energy resolution is assumed to be 1.6 keV over all energies in approximate agreement with measurements \cite{liu2014silicon}. To model pileup distortion in the silicon-strip detector, the nonparalyzable model is applied with a dead time of 35 ns \cite{liu2016count,knoll2010radiation}. The model agrees with the measured data for relatively low levels of pileup \cite{liu2016count}. Photoelectric effect and Compton scattering are modeled in the detector response function \cite{bornefalk2010photon}. The charge sharing effect is neglected in the silicon-strip detector because charge sharing is around 3\% in lower energy segments and is further reduced by the anticoincidence logic in the readout electronics \cite{xu2011validity,persson2014energy,ballabriga2013medipix3rx,koenig2013charge}.

The CZT detector model is originated from J. P. Schlomka \cite{schlomka2008experimental}, and has a maximum of 6 bins and a pixel size of 0.5 $\times$ 0.5 mm$^2$ \cite{8069518,6728030}. Although there are CZT detectors with a smaller pixel size and less pileup \cite{Kappler2010A}, it will result in severer charge sharing and we therefore do not expect the response function used here to be applicable for such systems. The thickness of the CZT detector is 3 mm with a 0.02 mm dead layer. The dead time is 40 ns for pileup distortion by applying the nonparalyzable model. The energy resolution is given by $\sigma= a_1+a_2\cdot E$, where $a_1$ is 1.61 keV and $a_2$ is 0.025 \cite{schlomka2008experimental}. In the detector response function, two Gaussian peaks are included to simulate the photoelectric effect and K-fluorescence, with one at the incident energy and the other at the energy reduced by the K-fluorescence energy. A step-like function is also included as background to simulate the charge sharing effect \cite{schlomka2008experimental}.

Pulse pileup can generate a distorted spectrum due to the stochastic arrival of photons and the dead time after counting each photon, which all photon-counting detectors suffer from. Overlapping pulses may be detected as a single count with a higher energy. To simulate pulse pileup in both detectors, the Kth-order delta pulse pileup model is applied to obtain the output distorted spectrum \cite{wang2011pulse}, and the multinomial distribution replaces the Poisson distribution for the total detected counts in each bin \cite{wang2011pulse}. Charge sharing is a significant effect in the CZT detector and results in incomplete charge collection, double counting and degraded image quality. If an incoming photon with an actual energy E$_0$ interacts between two neighboring pixels, the deposited energy in one pixel should always be larger than E$_0$/2 as a result of charge sharing \cite{liu2014silicon,xu2013energy}. We therefore assume that the spectral distribution of primary counts, i.e excluding charge-shared counts from neighboring pixels, is obtained by discarding the deposited energies below E$_0$/2.

The first FOM that we will use for optimizing the thresholds is the SDNR$^2$ of an optimally weighted image. We look at the beam incident on one single pixel to avoid considering noise correlations among pixels. Factoring in pulse pileup and charge sharing, and taking into account the optimal weighting factor $\bi w$ \cite{barrett2013foundations,bornefalk2011task}, SDNR$^2$ in the projection-based energy weighting method is given by \cite{barrett2013foundations}

\begin{eqnarray}
{\rm SDNR^2} &= \frac{(\bi w \Delta \bi g^T)^2}{\bi w(\bi K^{b}+\bi K^{t})\bi w^T} \nonumber \\
&= \Delta \bi g^T(\bi K^{b}+\bi K^{t})^{-1}\Delta \bi g
\label{eq:sdnr4}
\end{eqnarray}
where $\Delta \bi g=\bi g^b-\bi g^t$ is the difference between background counts $\bi g^b=(I_1^b,...,I_N^b)^T$ and target counts $\bi g^t=(I_1^t,...,I_N^t)^T$; $\bi K^b$ and $\bi K^t$ are the covariance matrices of $\bi g^b$ and $\bi g^t$ respectively; and $I_i$ is the expected count in bin $B_i$ with pileup considered, (i = 1,2,...,N). For the CZT detector, the charge sharing effect makes it more complicated to calculate SDNR$^2$ since the counts in adjacent pixels will be correlated. Several authors have proposed models for spatial noise correlations in photon-counting detectors \cite{Faby2016An,Rajbhandary2017Effect,Stierstorfer2017Modeling,Xu2014Cascaded}, but it is presently not known how these correlations are affected by pileup. In the present study we therefore avoid the need for modeling spatial noise correlations, by assuming that the target is small enough to cover only a single pixel so that the charge sharing counts in the target case come from the surrounding pixels which contain only background information. This allows us to assess the detection performance for small features in the presence of pileup without a model for how pileup affects correlations. The background bin counts $I_i^b$ are calculated by propagating the full background spectrum $f^{b,cs}$, including charge sharing counts, through the Kth-order delta pulse pileup model. In the target case, $I_i^t$ are calculated by propagating the spectrum $f^{t,cs}_* = f^{b,cs}+f^{t,ncs}-f^{b,ncs}$, through the Kth-order delta pulse pileup model. $f^{t,cs}_*$ is the sum of the background spectrum and a difference spectrum caused by the insertion of the target, where the difference spectrum does not include charge sharing counts from neighboring pixels, since these pixels are unaffected by the target.

The second FOM that we will use is the SNR$^2$ of a material specific image resulting from basis decomposition. Three bases are chosen as water, bone and gadolinium (Gd). Thus the attenuation of the target case could be written as $\int \mu_{\rm target} ds = A_{\rm water}\mu_{\rm water} + A_{\rm bone}\mu_{\rm bone} + A_{\rm Gd}\mu_{\rm Gd}$, where $A_{\rm water}$, $A_{\rm bone}$ and $A_{\rm Gd}$ are the line-integrals of the corresponding basis material. Based on the above mentioned spectra considering pileup, the CRLB is used to evaluate the basis noise and is calculated as the inverse matrix of the Fisher information matrix $\mathcal{F}$ \cite{wang2011pulse}. Thus the lower bound for the variance of the Gd basis is $\sigma^{2}_{A_{\rm Gd}} \ge {\rm CRLB}_{\rm Gd,Gd}$. Factoring in pileup and charge sharing, SNR$^2$ is given by
\begin {equation}
{\rm SNR^2} = \frac{A_{\rm Gd}^2}{\sigma^{2}_{A_{\rm Gd}}} \le  \frac{A_{\rm Gd}^2}{{\rm CRLB}_{\rm Gd,Gd}}
\label{eq:snr}
\end {equation}
In practice, the maximum likelihood method tends to give a variance $\sigma^{2}_{A_{\rm Gd}}$ close to that predicted by the CRLB (${\rm CRLB}_{\rm Gd,Gd}$) \cite{roessl2011sensitivity}.

The optimal thresholds for a photon-counting spectral CT system are to maximize SDNR$^2$ in Eq. \ref{eq:sdnr4} or SNR$^2$ in Eq. \ref{eq:snr}. To search for the optimal thresholds, the Global-Search method based on the interior-point algorithm implemented in MATLAB is used to find the multi-dimensional maximal SDNR$^2$ or SNR$^2$. Local maximal SDNR$^2$ or SNR$^2$ could be reached from various starting points generated through a scatter-search mechanism, where the largest local maximum is regarded as the global maximum. The corresponding thresholds are taken as the optimal thresholds. At the optimal SDNR$^2$, the difference of SDNR$^2$ over a small difference of thresholds is given by the Taylor series stopping at the second order derivative
\begin{eqnarray}
|\Delta {\rm SDNR^2}| &= |\nabla {\rm SDNR^2}(\bi T)^T \Delta \bi T + \frac{1}{2} \Delta \bi T^T \bi H(\bi T) \Delta \bi T + O(\Delta \bi T^2)| \nonumber \\
&= |\frac{1}{2} \Delta \bi T^T \bi P^T \bi \Lambda \bi P \Delta \bi T| \nonumber \\
&\leq \frac{1}{2} |\lambda|_{max}  \parallel \bi u \parallel^2
\label{eq:deltasdnr}
\end{eqnarray}
where $\bi T=(t_1,t_2,...,t_N)$ is the threshold placement vector; the derivative $\nabla {\rm SDNR^2}(\bi T) = 0$ at the optimal SDNR$^2$; $\bi H(\bi T) = \bi P^T \bi \Lambda \bi P$ is the eigendecomposition of the Hessian matrix; and $\bi P\Delta \bi T = \bi u$ is regarded as a new variable. When $|\Delta \bi T |$ is 1 keV, $|\bi u| = |\Delta \bi T|$ since $\bi P$ is an orthogonal matrix. Thus $|\Delta {\rm SDNR^2}|$ is less than $|\lambda|_{max}/2$, where $|\lambda|_{max}$ is the maximum absolute eigenvalue of Hessian matrix $\bi H(\bi T)$. Therefore, $|\lambda|_{max}/2$ theoretically reflects the decrease of optimal SDNR$^2$ when the distance between the selected thresholds and the optimal thresholds, measured as the vector norm between the two sets of thresholds, is 1 keV.

Both SDNR$^2$ and $|\lambda|_{max}$ are normalized by the corresponding ideal SDNR$^2$ calculated by covering the deposited energy range with 1 keV bins, and these quantities are denoted as relative SDNR$^2$ and relative decrease of SDNR$^2$ respectively. Similarly, SNR$^2$ is normalized by the ideal SNR$^2$ and denoted as relative SNR$^2$. Since we look at the relative SDNR$^2$ and the relative SNR$^2$ instead of the absolute values, these figures of merit are independent of the absolute photon flux. When the phantom size increases, they thus only show the effect of the change of spectrum shape, not the change in quantum noise level. Both SDNR$^2$ and SNR$^2$ give the upper limits of how good performance could be reached when considering pileup and charge sharing. Although the spectral distortion caused by the charge sharing effect could be corrected by post-acquisition correction algorithms \cite{Taguchi2018spatio,Christensen2017spectral,Taguchi2018SpatioenergeticCI}, the performances of SDNR$^2$ and SNR$^2$ could not be further improved. Thus the correction algorithms are not applied in this study. For each imaging target and each number of bins, Global-Search is run 20 times, which is enough to obtain stable optimal thresholds in real practice. The first threshold in the silicon-strip detector is set to 5 keV to discard electronic noise \cite{bornefalk2010photon}. In the CZT detector, the placement of the first threshold is optimized.

\section{Results}
\label{sect:result}
\begin{figure*} [!ht]
	\centering%
	\subfloat[D = 15 cm]{%
		\label{fig:plot115}
		\includegraphics[height = 5.5cm,width = 8cm]{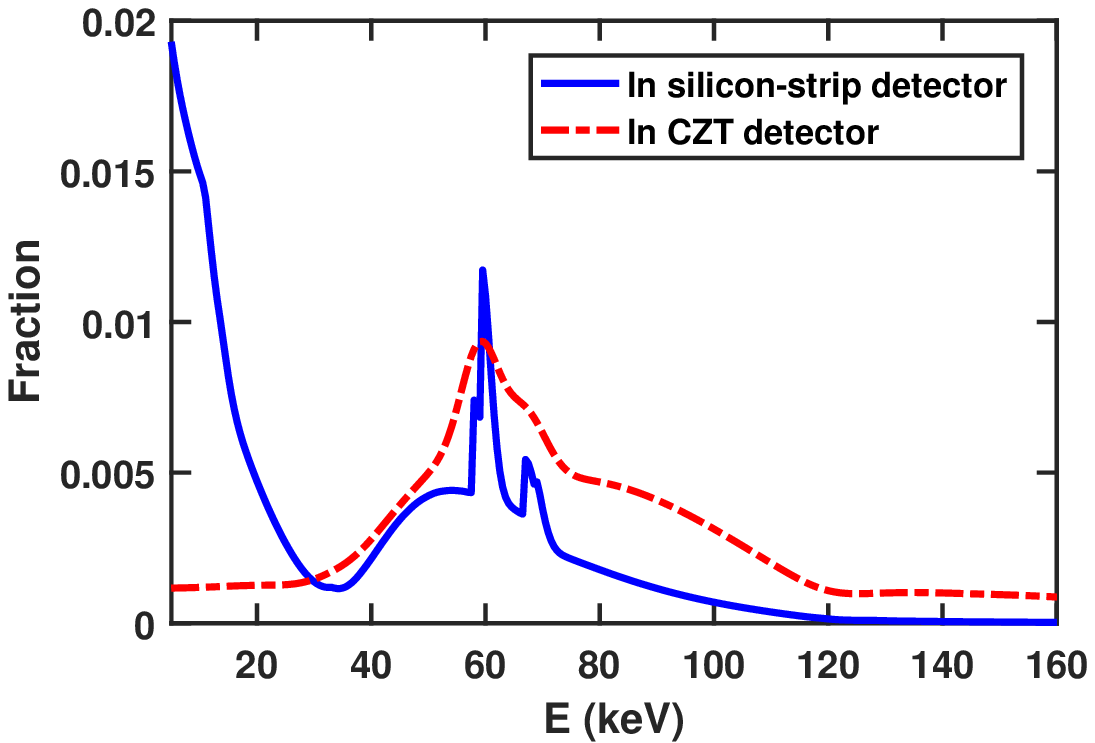}\hspace{0.5em}}%
	\subfloat[D = 50 cm]{%
		\label{fig:plot150}
		\includegraphics[height = 5.5cm,width = 8cm]{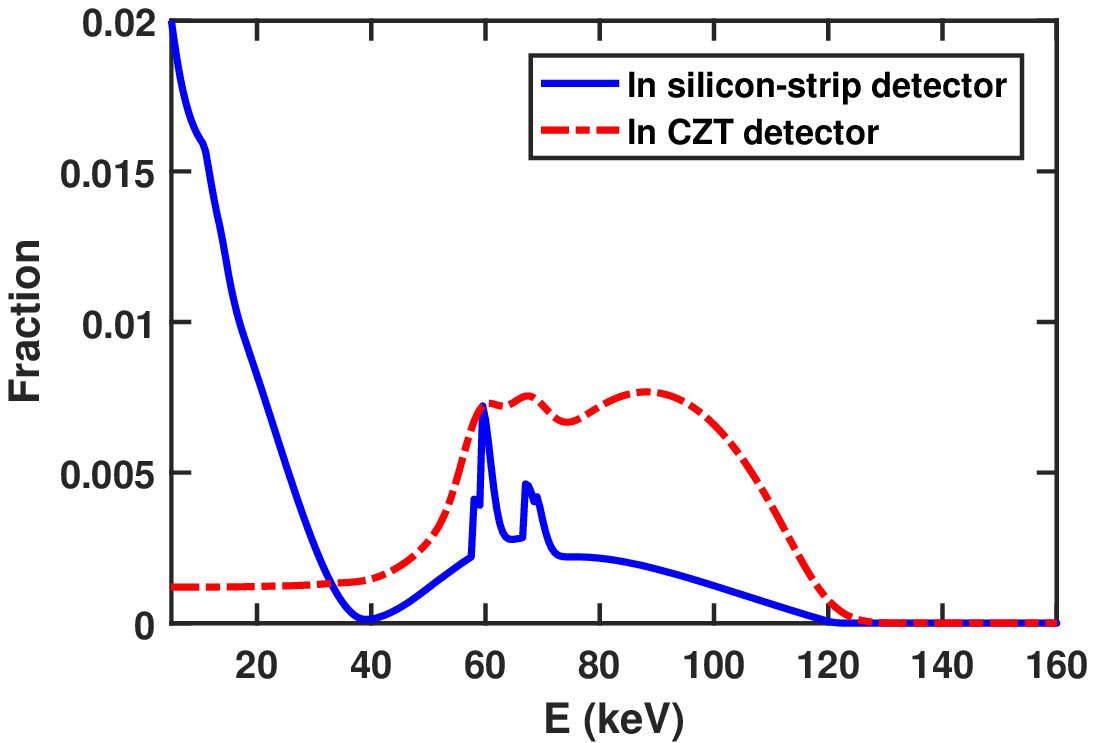}}
	\caption{Spectra of 10 mg/ml iodine in the silicon-strip detector and the CZT detector considering pileup. The y-axis "Fraction" refers to the fraction of photons that deposit energies in the detector after passing through the target and the phantom. The phantom thickness is (a) 15 cm and (b) 50 cm respectively.}
	\label{fig:plot1}
\end{figure*}

When the target is 10 mg/mL iodine and the phantom thickness D is 15 cm and 50 cm respectively, the spectra with pileup in the silicon-strip detector and the CZT detector are shown in Fig. \ref{fig:plot1}. With phantom thickness D = 15 cm, the fraction of the pileup counts above 120 kVp in Fig. \ref{fig:plot115} is about 0.6\% in the silicon-strip detector and 12.8\% in the CZT detector. In comparison, the fraction of pileup decreases to 0.02\% in the silicon-strip detector and 0.5\% in the CZT detector with a thicker phantom in Fig. \ref{fig:plot150}.

\subsection{SDNR$^2$ as a FOM}
\begin{figure}[!ht] 
	\centering
	\includegraphics[height = 7cm,width = 8cm]{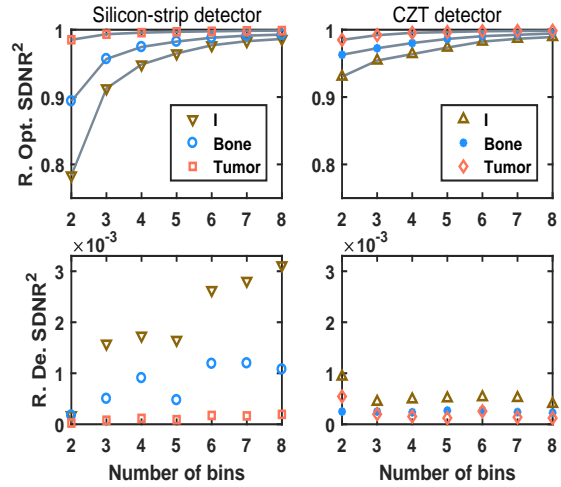}
	\caption{Relative optimal SDNR$^2$ and relative decrease of optimal SDNR$^2$ as a function of bin numbers in the silicon-strip detector and the CZT detector when considering pileup. The relative decrease of optimal SDNR$^2$ is given by $|\lambda|_{max}/2$ in Eq. \ref{eq:deltasdnr} normalized by the ideal SDNR$^2$, when the distance between the selected thresholds and the optimal thresholds is 1 keV.}
	\label{fig:plot2}
\end{figure}

When applying SDNR$^2$ as a FOM, to investigate the performance of various target materials in the threshold optimization, targets are set as 10 mg/ml iodine, bone or tumor respectively and the phantom thickness is kept 30 cm. Considering pileup in both detectors, figure \ref{fig:plot2} presents relative optimal SDNR$^2$ and the relative decrease of optimal SDNR$^2$ when the distance between the selected thresholds and the optimal thresholds is 1 keV, as a function of bin numbers and targets.

\begin{figure*}[!ht] 
	\centering
	\includegraphics[width = 15cm]{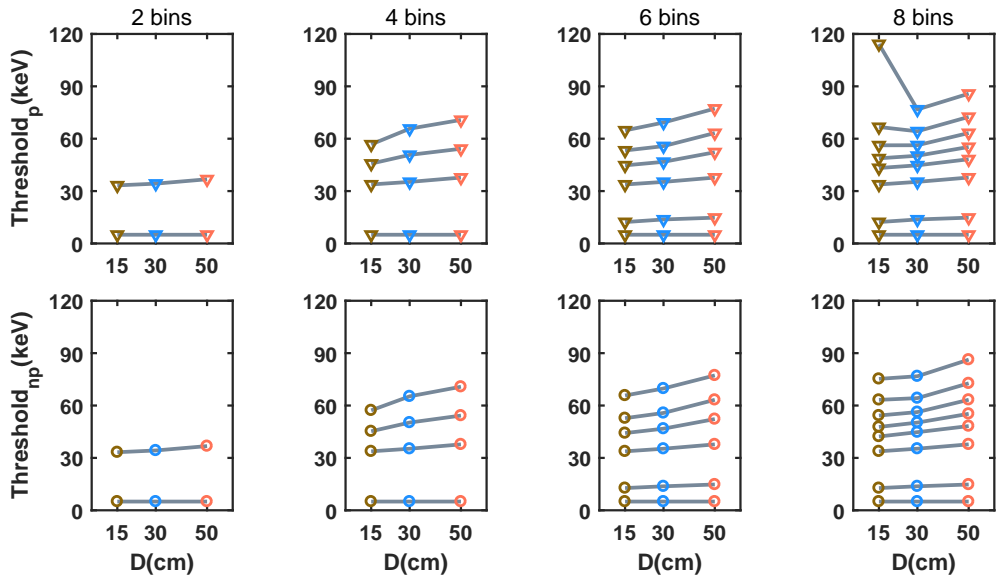}
	\caption{Optimal thresholds as a function of phantom thickness D in the silicon-strip detector when using SDNR$^2$ as a FOM. The upper row considers pileup and the lower row neglects pileup.}
	\label{fig:plot3si}
\end{figure*}

\begin{figure*}[!ht] 
	\centering
	\includegraphics[width = 15cm]{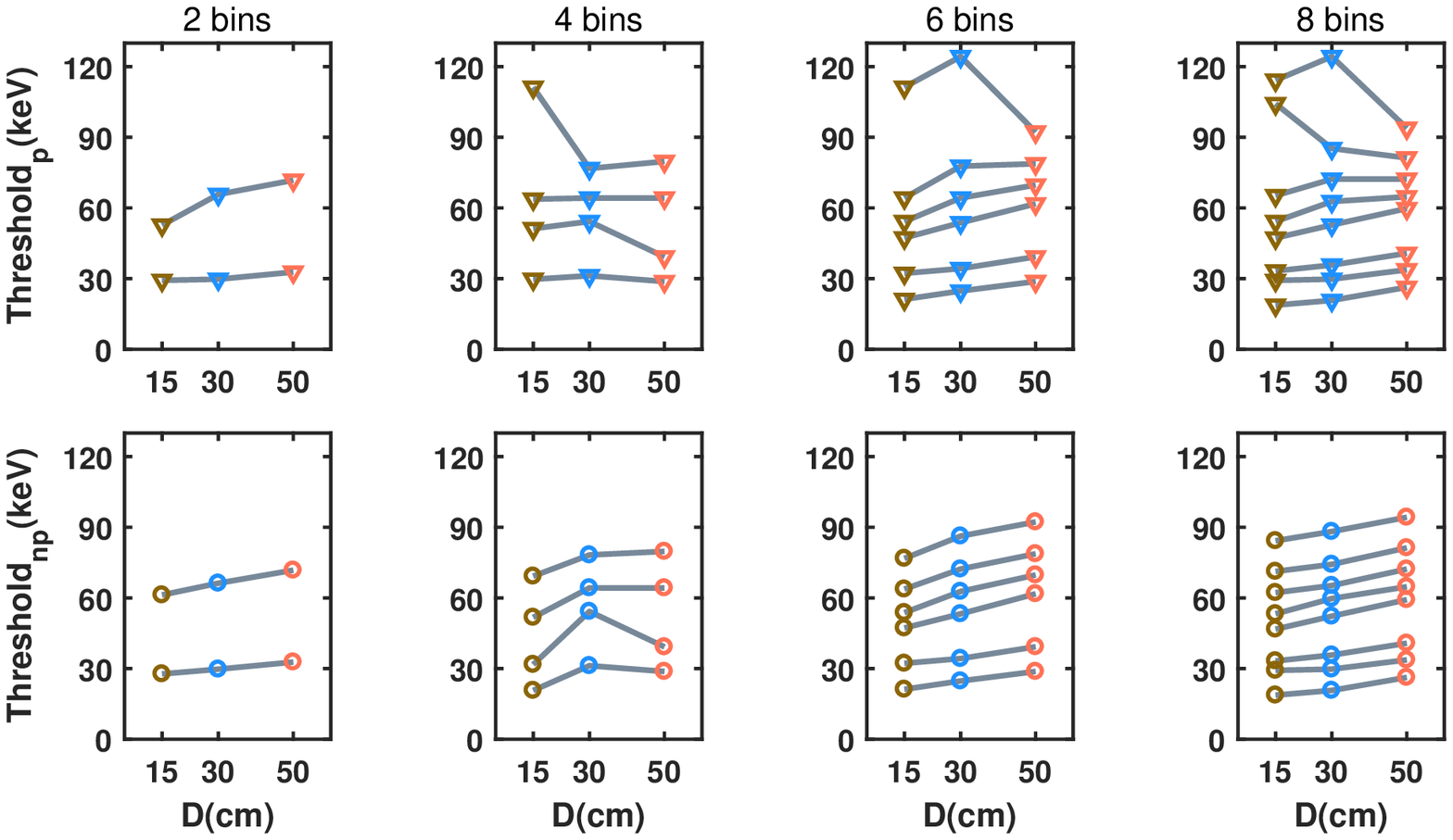}
	\caption{Optimal thresholds as a function of phantom thickness D in the CZT detector when using SDNR$^2$ as a FOM. The upper row considers pileup and the lower row neglects pileup.}
	\label{fig:plot3czt}
\end{figure*}

Phantom thickness is important to the optimal threshold placement. To investigate the dependence of optimal thresholds on phantom thickness, optimal thresholds with 2 to 8 bins are calculated when the target is 10 mg/mL iodine and the phantom thickness is changed from 15 cm to 30 cm and 50 cm respectively. The optimal thresholds for the silicon-strip detector and the CZT detector are shown in Fig. \ref{fig:plot3si} and \ref{fig:plot3czt}, factoring in pileup and no pileup separately.

\begin{figure*}[!ht] 
	\centering
	\includegraphics[width = 15cm]{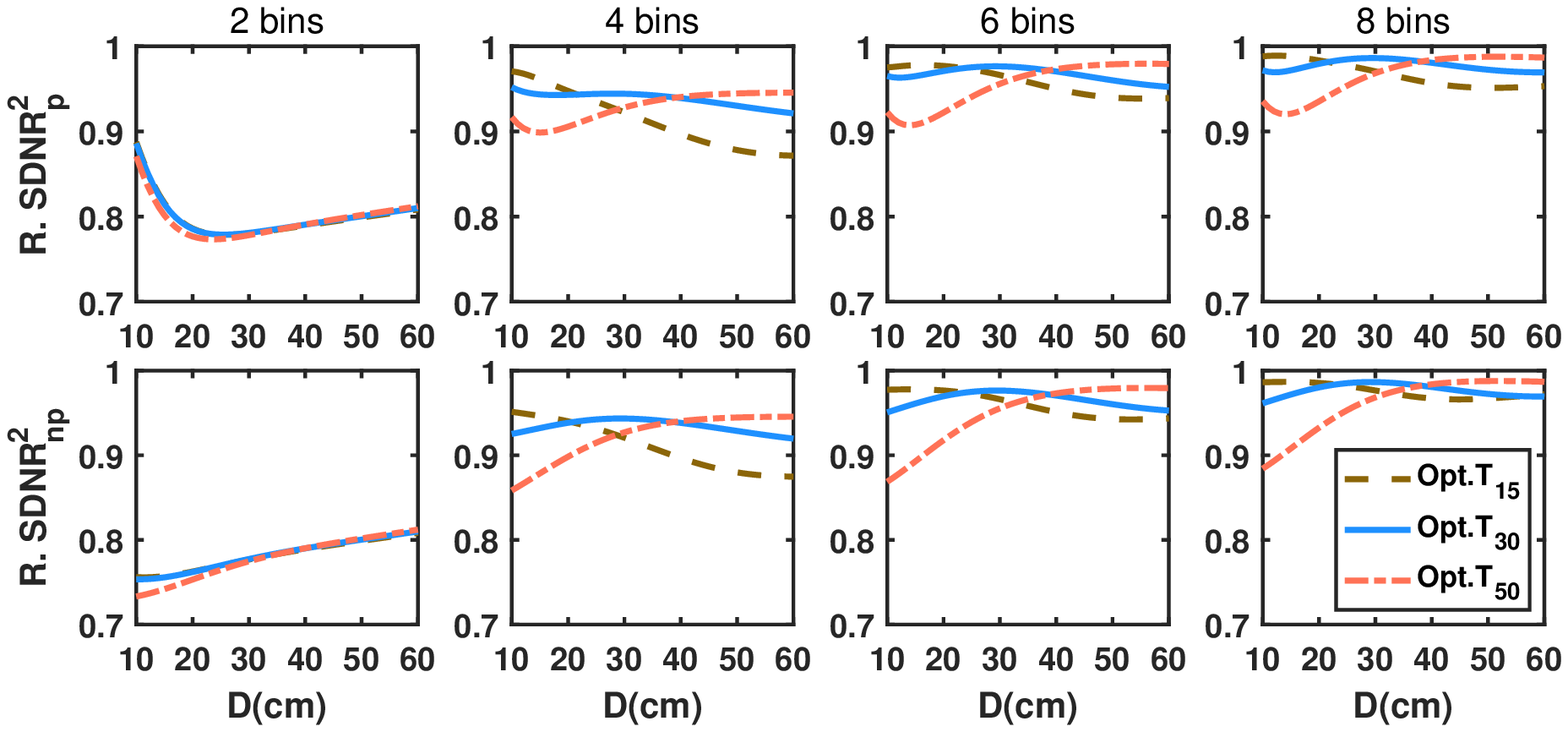}
	\caption{SDNR$^2$ as a function of phantom thickness in the silicon-strip detector when thresholds are fixed as the optimal thresholds for a 15 cm, 30 cm and 50 cm phantom in Fig. \ref{fig:plot3si}. The upper row considers pileup and the lower row neglects pileup.}
	\label{fig:plot4si}
\end{figure*}

\begin{figure*}[!ht] 
	\centering
	\includegraphics[width = 15cm]{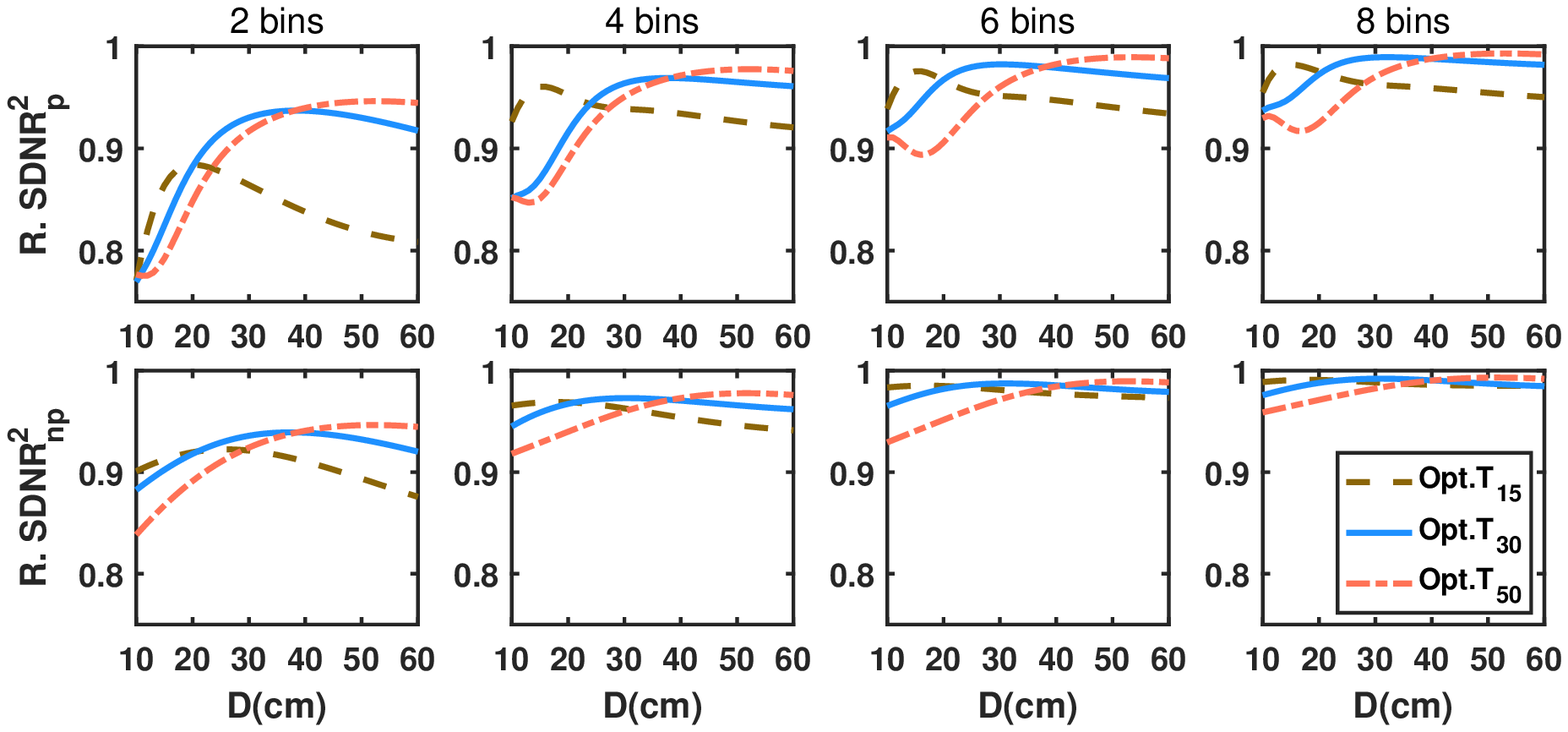}
	\caption{SDNR$^2$ as a function of phantom thickness in the CZT detector when thresholds are fixed as the optimal thresholds for a 15 cm, 30 cm and 50 cm phantom in Fig. \ref{fig:plot3czt}. The upper row considers pileup and the lower row neglects pileup.}
	\label{fig:plot4czt}
\end{figure*}

\begin{figure*}[!ht] 
	\centering
	\includegraphics[width = 15cm]{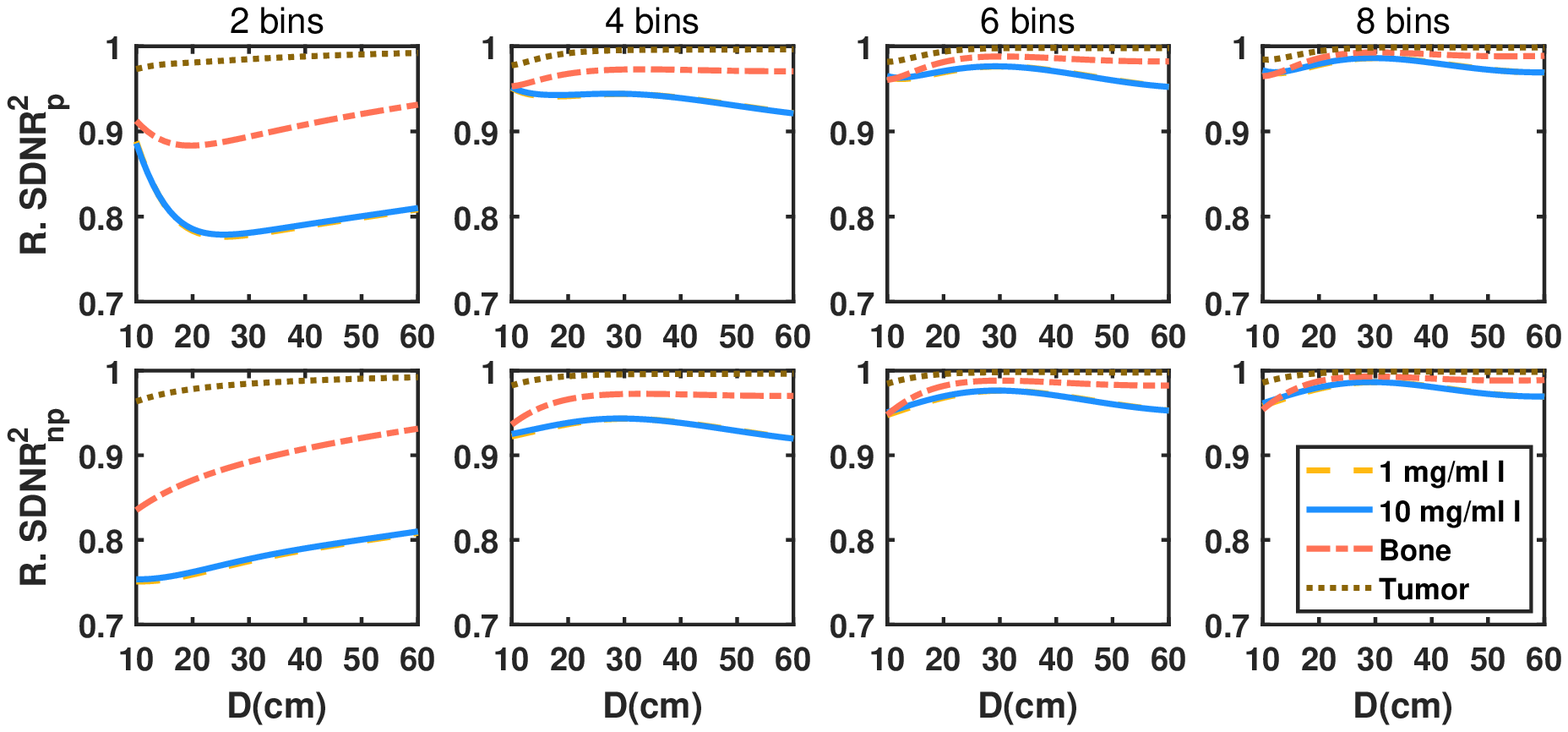}
	\caption{SDNR$^2$ as a function of phantom thickness and target materials in the silicon-strip detector when thresholds are fixed as the optimal thresholds for a 30 cm phantom and iodine target in Fig. \ref{fig:plot3si}. The upper row considers pileup and the lower row neglects pileup.}
	\label{fig:plot5si}
\end{figure*}

\begin{figure*}[!ht] 
	\centering
	\includegraphics[width = 15cm]{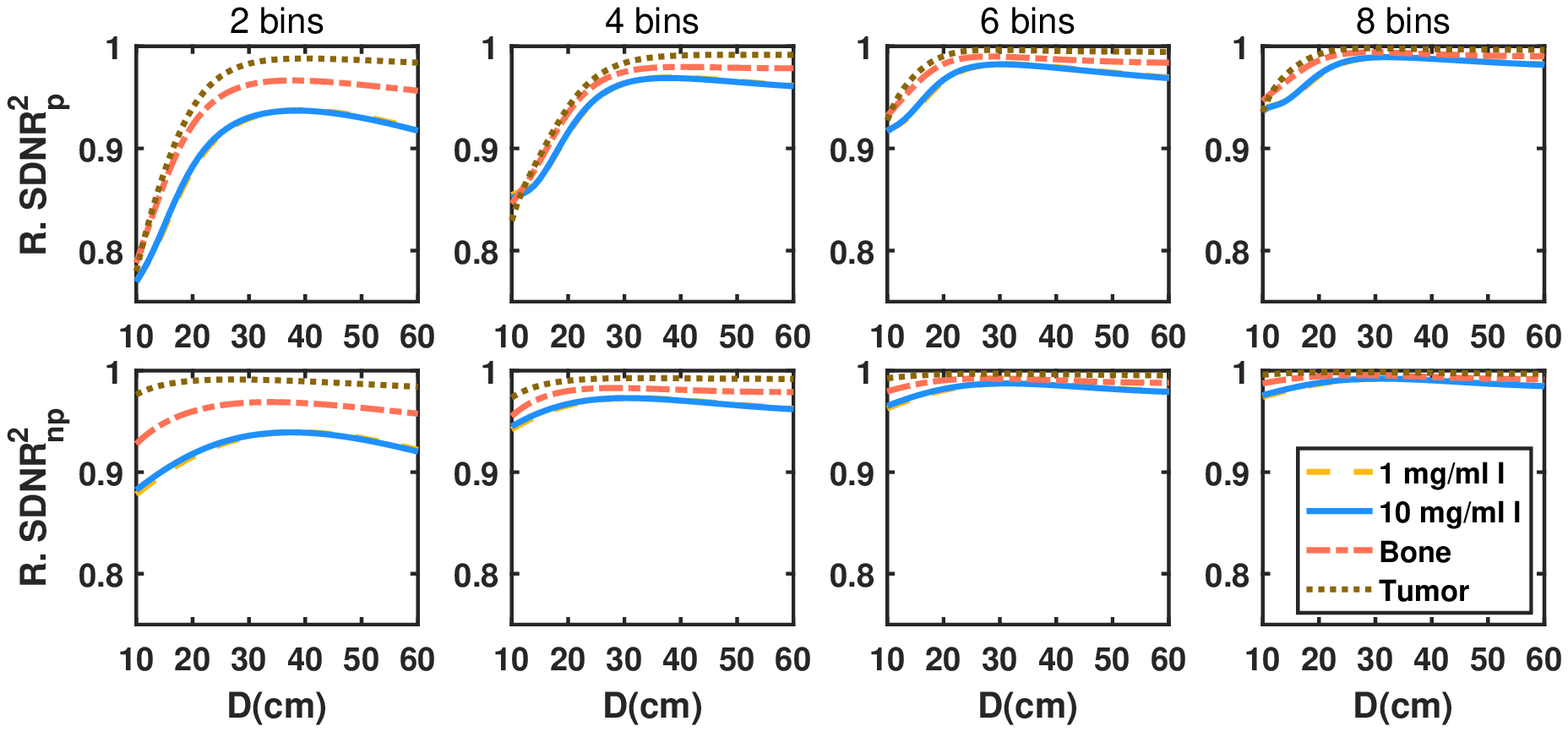}
	\caption{SDNR$^2$ as a function of phantom thickness and target materials in the CZT detector when thresholds are fixed as the optimal thresholds for a 30 cm phantom and iodine target in Fig. \ref{fig:plot3czt}. The upper row considers pileup and the lower row neglects pileup.}
	\label{fig:plot5czt}
\end{figure*}

To investigate the robustness of optimal thresholds and to what extent one single set of energy thresholds could be applied, the variation of SDNR$^2$ with phantom thicknesses and target materials is examined when the thresholds are fixed as the optimal thresholds for iodine. When the thresholds are fixed as the optimal thresholds for phantom 15 cm, 30 cm and 50 cm in Fig. \ref{fig:plot3si} and \ref{fig:plot3czt}, relative SDNR$^2$ as a function of phantom thickness is shown in Fig. \ref{fig:plot4si} in silicon-strip detector and in Fig. \ref{fig:plot4czt} in CZT detector. When the thresholds are fixed as the optimal thresholds for phantom 30 cm only, relative SDNR$^2$ for a target of 1 mg/ml iodine, 10 mg/ml iodine, bone or tumor is shown in Fig. \ref{fig:plot5si} and \ref{fig:plot5czt} with the variation of phantom thickness.

\subsection{SNR$^2$ as a FOM}

\begin{figure*}[!ht] 
	\centering
	\includegraphics[width = 15cm]{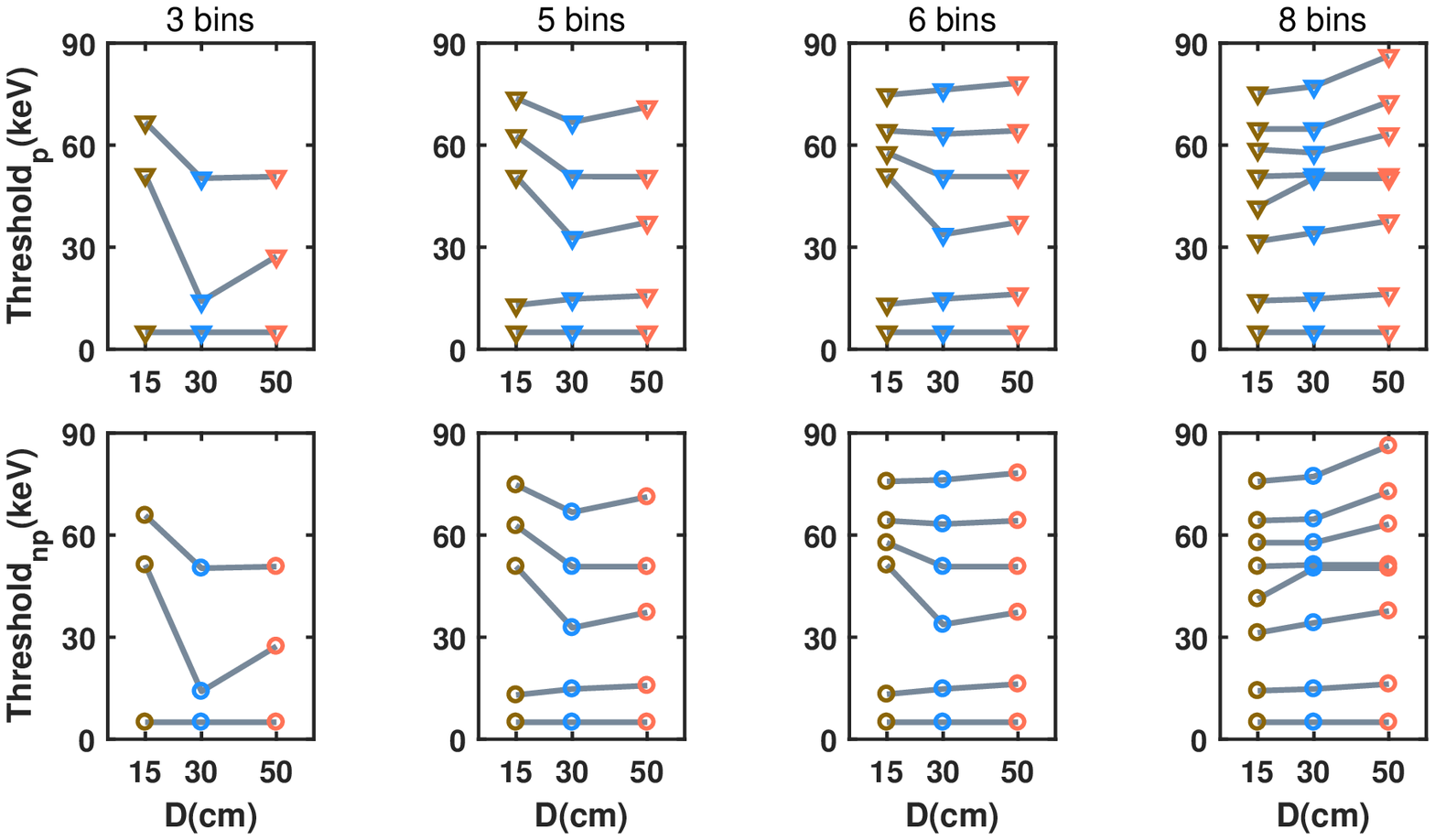}
	\caption{Optimal thresholds as a function of phantom thickness D in the silicon-strip detector when using SNR$^2$ as a FOM. The upper row considers pileup and the lower row neglects pileup.}
	\label{fig:plot6si}
\end{figure*}

\begin{figure*}[!ht] 
	\centering
	\includegraphics[width = 15cm]{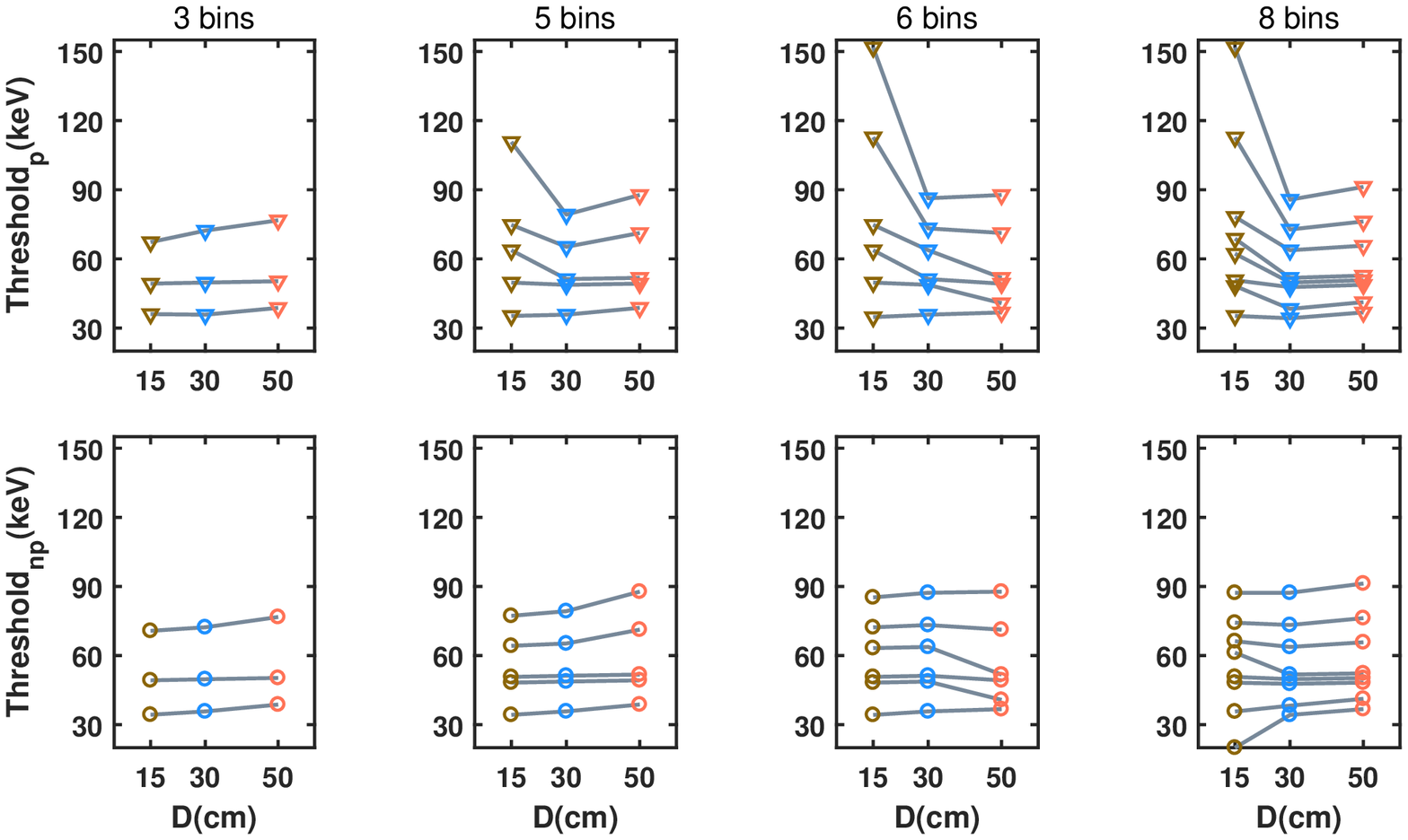}
	\caption{Optimal thresholds as a function of phantom thickness D in the CZT detector when using SNR$^2$ as a FOM. The upper row considers pileup and the lower row neglects pileup.}
	\label{fig:plot6czt}
\end{figure*}

When using SNR$^2$ in the Gd basis image as a FOM, the target is set to 1 mg/ml Gd and optimal thresholds are calculated with increasing phantom thicknesses. The results are shown in Fig. \ref{fig:plot6si} for the silicon-strip detector and in Fig. \ref{fig:plot6czt} for the CZT detector.

\begin{figure*}[!ht] 
	\centering
	\includegraphics[width = 15cm]{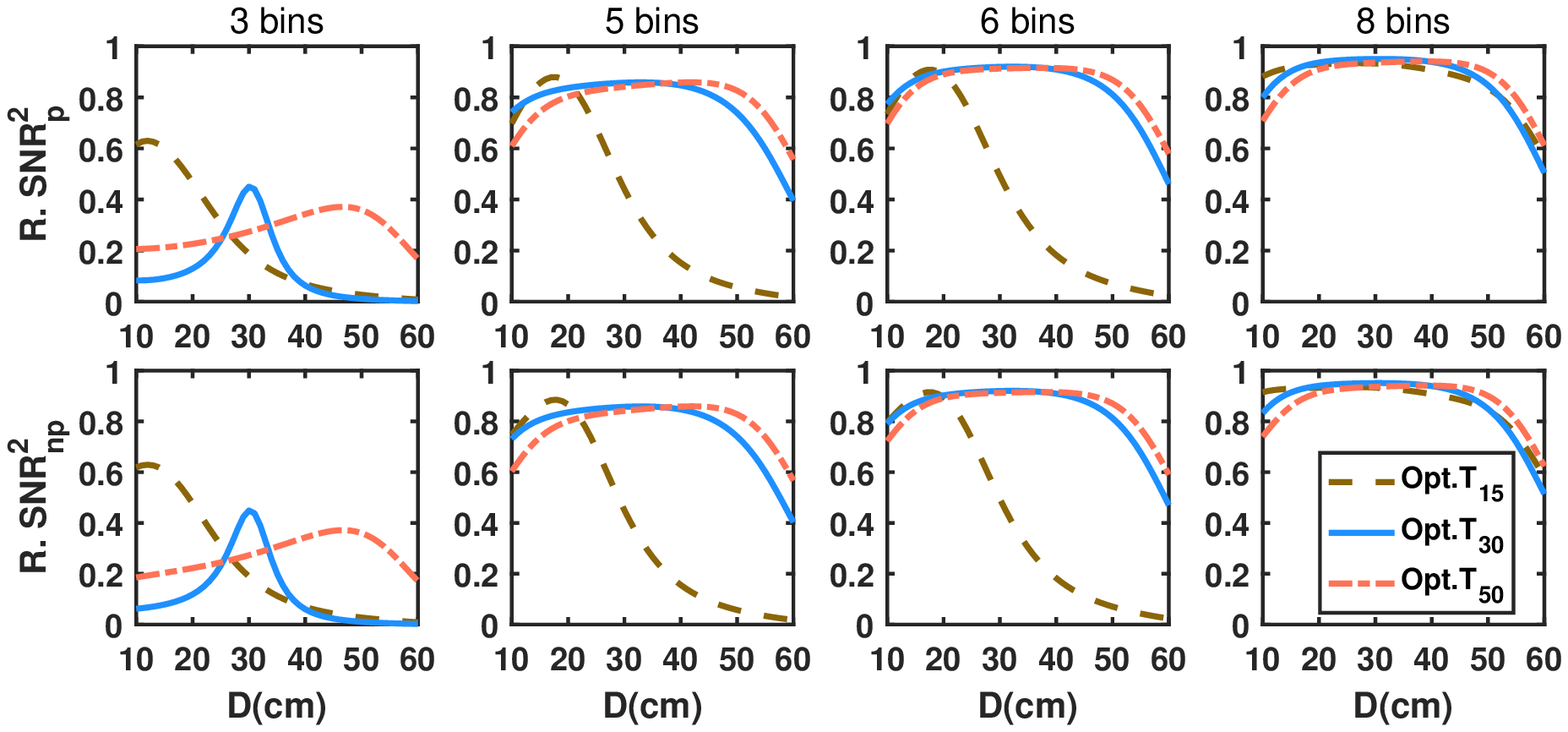}
	\caption{SNR$^2$ as a function of phantom thickness in the silicon-strip detector when thresholds are fixed as the optimal thresholds for a 15 cm, 30 cm and 50 cm phantom in Fig. \ref{fig:plot6si}. The upper row considers pileup and the lower row neglects pileup.}
	\label{fig:plot7si}
\end{figure*}

\begin{figure*}[!ht] 
	\centering
	\includegraphics[width = 15cm]{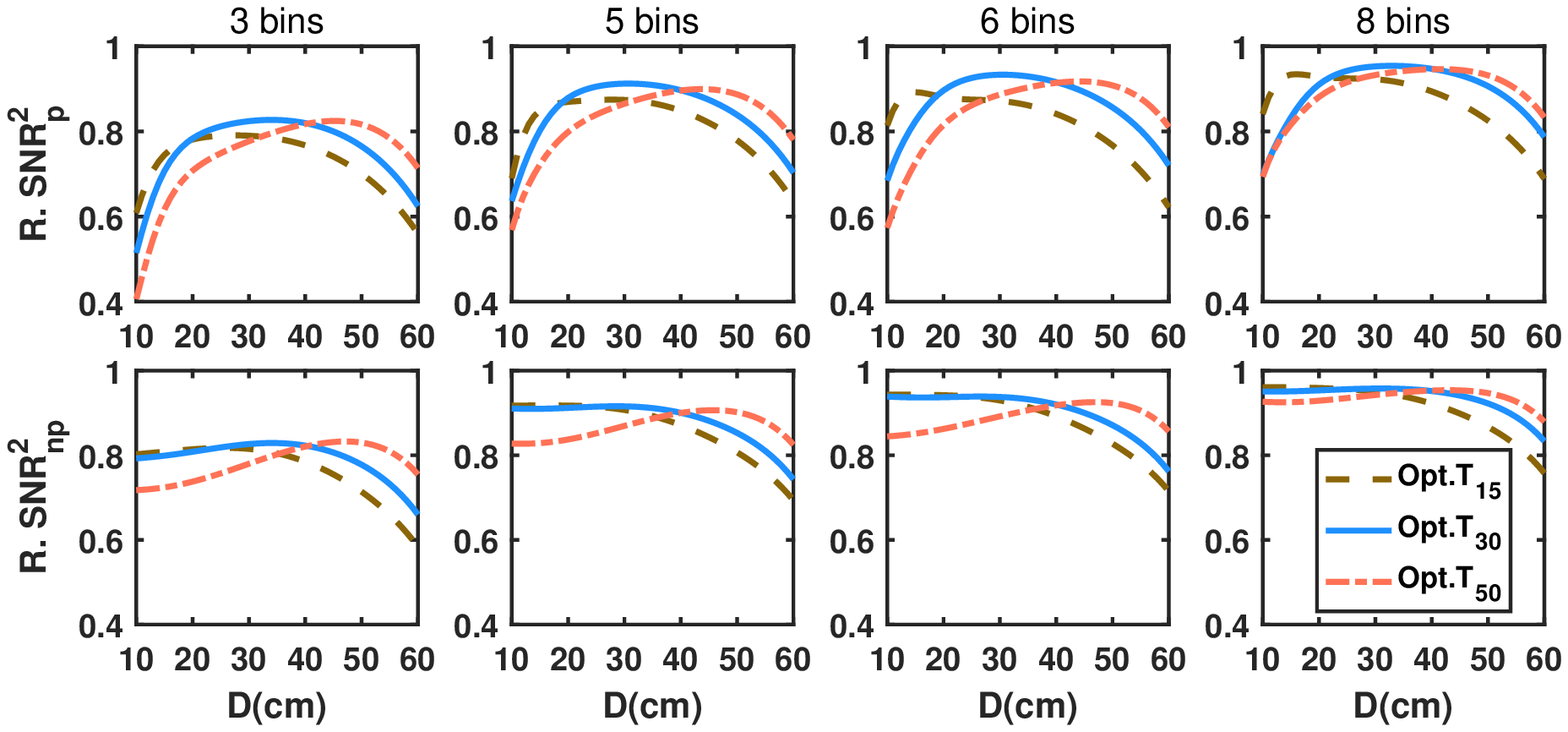}
	\caption{SDNR$^2$ as a function of phantom thickness in the CZT detector when thresholds are fixed as the optimal thresholds for a 15 cm, 30 cm and 50 cm phantom in Fig. \ref{fig:plot6czt}. The upper row considers pileup and the lower row neglects pileup.}
	\label{fig:plot7czt}
\end{figure*}

Similarly, to investigate the robustness of optimal thresholds when using SNR$^2$ as the FOM, the thresholds are fixed as the optimal thresholds for phantom 15 cm, 30 cm and 50 cm in Fig. \ref{fig:plot6si} and \ref{fig:plot6czt}, relative SNR$^2$ as a function of phantom thickness is shown in Fig. \ref{fig:plot7si} in the silicon-strip detector and in Fig. \ref{fig:plot7czt} in the CZT detector.

\section{Discussion}
\label{sect:discussion}
Optimal SDNR$^2$ or SNR$^2$ and the placement of optimal thresholds are highly related to the detected spectrum, which is influenced by detectors, target compositions and phantom thicknesses. In this study the silicon-strip detector and the CZT detector are modeled as two example detector systems to examine the robustness of optimal thresholds, and the study should not be seen as a performance comparison of the two systems since the results depend on specific assumptions about the parameters of these systems. The CZT detector suffers a lot from pulse pileup and charge sharing has worse energy resolution (see Fig. \ref{fig:plot1}). The silicon-strip detector has better energy resolution but yields a higher fraction of Compton scattering (see Fig. \ref{fig:plot1}). For each detector, the target compositions and the surrounding attenuation are two main influencing parameters in optimizing energy thresholds, and two concerns in examining the robustness of optimal thresholds.

Targets in clinical applications could be iodine contrast agent, tumor, bone, or a mixture of these objects. For both detectors with these targets, the relative optimal SDNR$^2$ increases with the number of bins (see Fig. \ref{fig:plot2}). For the silicon-strip detector, it's effective to improve optimal SDNR$^2$ by increasing the number of bins from 2 to 6 or 8 when keeping phantom thickness unchanged. For the CZT detector, on the other hand, the improvement obtained by adding more bins is smaller. The reason for this is that the CZT detector has deteriorated energy resolution, and a small number of bins are sufficient to capture the remaining energy information. In Fig. \ref{fig:plot2}, for an iodine target and 2 bins, the loss of optimal SDNR$^2$ reaches 20\% in the silicon-strip detector and about 7\% in the CZT detector. With 6 to 8 bins, the loss of optimal SDNR$^2$ is about 1\% to 2\% in both detectors. When the target is only tumor, setting more bins does not considerably further increase SDNR$^2$ in both detectors because photons arriving at the detector contain less energy attenuation information. For both detectors, it's difficult to predict how the relative decrease of optimal SDNR$^2$ for a threshold error with norm $|\Delta \bi T|$ = 1 keV should vary with increasing bin numbers. Given that SDNR$^2$ is not affected by the quantum noise level, so the discontinuity of the relative decrease of optimal SDNR$^2$ in Fig. \ref{fig:plot2} is not due to the impact of noise. Since the relative decrease of optimal SDNR$^2$ is still less than 0.4\% of the corresponding optimal SDNR$^2$ (see Fig. \ref{fig:plot2}), it indicates that each optimal SDNR$^2$ is stable enough to threshold fluctuations regardless of bin numbers. Thus, it is not necessary to set the thresholds with very high precision in applications.

The phantom thickness plays an important role in threshold optimization because a larger thickness filters the detected spectrum more and changes the fraction of pulse pileup (see Fig. \ref{fig:plot1}). Previous studies have shown that the optimal threshold energies increase with phantom thickness because of the beam hardening effect \cite{shikhaliev2009projection,wang2009optimal,roessl2006optimal,roessl2011sensitivity}, which is also demonstrated in Fig. \ref{fig:plot3si} and \ref{fig:plot3czt} when pileup is neglected in both detectors and SDNR$^2$ is used as the FOM. However, with pileup considered or SNR$^2$ used as the FOM, the optimal threshold energies in both detectors don't always increase or even decrease with increasing phantom thicknesses (see Fig. \ref{fig:plot3si} to \ref{fig:plot3czt} and Fig. \ref{fig:plot6si} to \ref{fig:plot6czt}). When pileup is included in the CZT detector, with a small phantom like 15 cm, pileup is severe and some highest optimal thresholds lie around or above 120 keV in Fig. \ref{fig:plot3czt} and \ref{fig:plot6czt}, indicating that the pileup counts above 120 kVp in Fig. \ref{fig:plot115} contain useful attenuation information. With a larger phantom, the pileup effect is weakened because more photons are attenuated away and less counts pile up. For both detectors, when the phantom is 50 cm, the optimal thresholds considering pileup are nearly the same as those neglecting pileup (see Fig. \ref{fig:plot3si} to \ref{fig:plot3czt} and Fig. \ref{fig:plot6si} to \ref{fig:plot6czt}). For the silicon-strip detector, the optimal threshold energies are not considerably changed by pileup no matter what FOM is evaluated (see Fig. \ref{fig:plot3si} and \ref{fig:plot6si}), because the output spectrum in the silicon-strip detector is not severely distorted by pileup (see Fig. \ref{fig:plot1}). When SNR$^2$ is the FOM in the silicon-strip detector, the optimal threshold energies increase with phantom thicknesses from 30 cm to 50 cm (see Fig. \ref{fig:plot6si}). In the silicon-strip detector, one optimal threshold locates at around 30 keV to separate Compton scattering counts with both FOMs evaluated (see Fig. \ref{fig:plot3si} and \ref{fig:plot6si}). For the CZT detector, the first optimal threshold is always higher than 20 keV regardless of bins in Fig. \ref{fig:plot3czt} and \ref{fig:plot6czt}, indicating that charge sharing counts deteriorate SDNR$^2$ or SNR$^2$ and should be discarded.

The close bracketing effect is also demonstrated when SNR$^2$ is used as the FOM in both detectors. However, the results in Fig. \ref{fig:plot6si} and \ref{fig:plot6czt} show that the close bracketing effect does not always occur unless enough bin numbers and less pileup, and sometimes with only one threshold around the K-edge of Gd (50.2 keV) optimal SNR$^2$ could be reached as well. The results are in agreement with the previous studies by Roessel et al., where the close bracketing effect was reported not to be very strong and not to occur for a range of phantom thicknesses \cite{roessl2011sensitivity}. In the silicon-strip detector, from 3 bins to 6 bins, only one threshold is optimized closely after the K-edge of Gd; and with 8 bins, 2 optimal thresholds closely bracket the K-edge when the phantom thickness is 30 cm to 50 cm (see Fig. \ref{fig:plot6si}). For the CZT detector in Fig. \ref{fig:plot6czt}, with 3 bins, only one optimal threshold locates before the K-edge of Gd because the number of bins is not enough to closely bracket the K-edge. From 5 to 6 bins in the CZT detector in Fig. \ref{fig:plot6czt}, if pileup is not severe (see D = 30 cm to 50 cm when including pileup or the no pileup case), 2 optimal thresholds closely bracket the K-edge of Gd, but otherwise only one optimal threshold locates before the K-edge. With 8 bins in the CZT detector, 2 or 3 optimal thresholds closely bracket the K-edge of Gd. Therefore it indicates that the phantom thickness has an important effect on the close bracketing effect and the optimal threshold energies in K-edge imaging as well.


In clinical applications, thresholds are set before exposure, and it is thus important to select the number of bins and the threshold energies to yield near-optimal SDNR$^2$ or SNR$^2$ over various attenuation and target materials. In both detectors the robustness of optimal thresholds is not sensitive to target materials but on the other hand sensitive to phantom thicknesses (see Fig. \ref{fig:plot4si} to \ref{fig:plot5czt}, and Fig. \ref{fig:plot7si} to \ref{fig:plot7czt}). For the silicon-strip detector, when SDNR$^2$ is the FOM, setting the optimal thresholds for a 30 cm phantom yields the largest mean and minimum SDNR$^2$ across different bin configurations compared with 15 and 50 cm phantoms, and the decrease of SDNR$^2$ is within 5\% with 6 to 8 bins regardless of pileup (see Fig. \ref{fig:plot4si}). When SNR$^2$ is the FOM, setting the optimal thresholds for a 50 cm phantom yields the most robust SNR$^2$ across different bin configurations in the silicon-strip detector and with increasing bin numbers, setting the optimal thresholds for a 30 cm phantom shows as robust performance as that of setting the optimal thresholds for a 50 cm phantom (see Fig. \ref{fig:plot7si}). Therefore, having more bins benefits the robustness of optimal thresholds in the silicon-strip detector but even with 6 to 8 bins the maximum decrease of SNR$^2$ is about 40\% when fixing the optimal thresholds for a 50 cm phantom regardless of pileup. For the CZT detector applying both FOMs, setting optimal thresholds for a 30 cm phantom yields the best SDNR$^2$ or SNR$^2$ across different bin configurations regardless of pileup (see Fig. \ref{fig:plot4czt} and \ref{fig:plot7czt}). However, SDNR$^2$ or SNR$^2$ as a function of phantom thicknesses decreases more when pileup is included in the CZT detector. In Fig. \ref{fig:plot4czt}, the maximum loss of SDNR$^2$ with 6 to 8 bins is 8\% when pileup is included, which is larger than the 4\% that is obtained when pileup is excluded. And in Fig. \ref{fig:plot7czt}, the maximum loss of SNR$^2$ with 6 to 8 bins is 30\% when fixing the optimal thresholds for a 30 cm phantom and including pileup, which is larger than the 25\% that is obtained when pileup is excluded. Technically, it's hard to fix one set of thresholds in both detectors in terms of the variation of attenuation when pileup is considered. According to Fig. \ref{fig:plot4si} to \ref{fig:plot4czt} and Fig. \ref{fig:plot7si} to \ref{fig:plot7czt}, if this decrease of SDNR$^2$ or SNR$^2$ is tolerable, setting thresholds for a 30 cm phantom approximately yields satisfactory SDNR$^2$ or SNR$^2$ with 6 to 8 bins in both detectors.

For the silicon-strip detector with 2 or 3 bins, the lowest threshold is fixed at 5 keV and the only one or two thresholds remain to be optimized (see Fig. \ref{fig:plot3si} and \ref{fig:plot6si}). However, optimizing just one or two thresholds is not enough to get optimal SDNR$^2$ or SNR$^2$ at a specific thickness (see Fig. \ref{fig:plot4si} and \ref{fig:plot7si}). The decrease of relative SDNR$^2$ from a 10 cm phantom to a 30 cm phantom with 2 bins in Fig. \ref{fig:plot4si} is caused by the fact that pileup degrades the ideal SDNR$^2$ more than the SDNR$^2$ of the simulated system. Figure \ref{fig:plot5si} and \ref{fig:plot5czt} demonstrate the robustness of optimal thresholds for a 30 cm phantom regardless of the variation of target materials when using SDNR$^2$ as the FOM. The relative SDNR$^2$ of bone and tumor is larger than that of iodine in both detectors when using the optimal thresholds for iodine. The plots of 1 mg/ml iodine and 10 mg/ml iodine almost overlap in Fig. \ref{fig:plot5si} and \ref{fig:plot5czt}, and we have examined that the optimal thresholds for various iodine concentrations within 5 mg/ml to 20 mg/ml are nearly the same.



In the current research, the counts scattered by the phantom is not considered. Also, it remains unknown how to model spatial noise correlations with pileup included and how to take them into account in silicon-strip detectors, and for this purpose we have chosen to use figures of merit (the SDNR$^2$ for an object covering a single pixel, and the SNR$^2$ in a basis image) that can be computed without modeling spatial noise correlations. A tradeoff between higher SDNR$^2$ or SNR$^2$ and less data needs to be considered to select suitable bin numbers. In Fig. \ref{fig:plot4si} and \ref{fig:plot4czt}, with 2 energy bins, up to 23\% of SDNR$^2$ is lost in both detectors; and in Fig. \ref{fig:plot7si} and \ref{fig:plot7czt}, with 3 bins, the maximum loss of SNR$^2$ is about 80\% when fixing the optimal thresholds for a 50 cm phantom in the silicon-strip detector and 50\% when fixing the optimal thresholds for a 30 cm phantom in the CZT detector. Thus it's better to have more than 3 bins. The more bins, the higher SDNR$^2$ or SNR$^2$. Conversely, when SDNR$^2$ or SNR$^2$ is close to the corresponding ideal value, further increasing bin numbers will not improve SDNR$^2$ or SNR$^2$ considerably but increase the amount of data. Based on that, using around 6 bins in both detectors helps reduce data and retain near-optimal SDNR$^2$ at the same time; and using about 8 bins in both detectors helps keep stable and near-optimal SNR$^2$.

\section{Conclusion}
\label{sect:conclusion}
In this work, the placement and robustness of optimal energy thresholds have been studied using SDNR$^2$ and SNR$^2$ as the FOM in a silicon-strip detector and a CZT detector, factoring in pileup and imperfect energy response in both detectors. Pileup has an important effect on the placement and robustness of optimal thresholds. Setting optimal thresholds for a 30 cm phantom approximately yields satisfactory SDNR$^2$ or SNR$^2$ in both detectors regardless of target materials and phantom attenuation. When fixing the optimal thresholds for a 30 cm phantom and considering pileup, with 6 to 8 bins, the maximum loss of SDNR$^2$ and SNR$^2$ is 5\% and 50\% in the silicon-strip detector respectively, and 8\% and 30\% in the CZT detector. And with 2 to 3 bins, up to 23\% of SDNR$^2$ could be lost in both detectors and the maximum loss of SNR$^2$ is about 80\% when fixing the optimal thresholds for a 50 cm phantom in the silicon-strip detector and 50\% when fixing the optimal thresholds for a 30 cm phantom in the CZT detector. Having more than 3 bins reduces the need for changing the thresholds depending on target materials and phantom thicknesses. There is a tradeoff to select bin numbers between higher SDNR$^2$ or SNR$^2$ and less data, with around 6 or 8 bins giving near-optimal SDNR$^2$ or SNR$^2$ without generating unnecessarily large amounts of data.


\ack
This study was supported by the China Scholarship Council (CSC). Fredrik Gr{\"o}nberg and Mats Persson have financial interests in Prismatic Senors AB.

\section*{References}
\bibliography{report}   
\bibliographystyle{jphysicsB}   
\bibliographystyle{harvard}

\end{spacing}
\end{document}